\documentclass[epjST,groupedaddress,floatfix]{svjour}

\usepackage{graphicx}
\usepackage{bm}
\usepackage{slashed}
\usepackage[utf8]{inputenc}
\usepackage{amsfonts}
\usepackage{amsmath}
\usepackage{amssymb}
\usepackage{graphicx,color}
\usepackage[colorlinks,linkcolor=blue,citecolor=green]{hyperref}
\usepackage{wrapfig}
\usepackage{cite} 
\newcommand{\dd}{\textrm{\,d}}

\renewcommand\thefigure{\thesection-\arabic{figure}}
\makeatletter
\@addtoreset{figure}{section}
\makeatother
\newcommand{\Lagr}{\mathcal{L}}

\begin{document}
\title{\Large \bf Estimating Compressibility from Maximal-mass Compact Star Observations}
\author{Gergely G. Barnaf\"oldi\inst{1}\fnmsep\thanks{\email{barnafoldi.gergely@wigner.hu}} \and P\'eter P\'osfay\inst{1}\and Bal\'azs E. Szigeti\inst{1,2}\and Antal Jakov\'ac\inst{2}}
\institute{Wigner Research Centre for Physics, P.O. Box, H-1525 Budapest, Hungary \and Institute of Physics, E\"otv\"os University, 1/A P\'azm\'any P. Stny., H-1117 Budapest, Hungary}

\date{\today}
\abstract{We investigated recent observation data of pulsar masses of PSR J0740$+$6620, PSR J0348$+$0432, and PSR J1614$-$2230 based on the extended $\sigma$-$\omega$ model. We assumed that these pulsars are maximal mass compact star, which suggest that the {\sl core approximation} can be applied. Using the linear relations between the microscopic and macroscopic parameters of neutron stars suggested by this model, we estimated the values of the nucleon Landau mass and nuclear compressibility $m_{L} = 776.0^{+38.5}_{-84.9}$~MeV and $K=242.7^{+57.2 }_{-28.0}$, respectively.} 

\maketitle

\section{Introduction}
\label{sec:sec1}

Pulsar's observables depend on the properties of the inner super-dense nuclear matter. Variation of the nature of the microscopic interactions and their parameters in the underlying nuclear theory can modify the magnitude of the mass and radius of a compact star. The inverse direction of this problem is to estimate nuclear parameters from neutron star observables, which is challenging due to the {\sl masquarade problem}: even large variation of the nuclear parameters and the change of the interaction terms result in stars with very similar macroscopic observable properties.

The effect of different interaction terms and the tuning of the nuclear parameter values in the Lagrangian has been presented so far for the case of the extended $\sigma$-$\omega$ model~\cite{Posfay:2020}. Linear dependence of the maximal mass and the corresponding radius parameters on the Landau mass $m_L$, compressibility $K$, and nuclear asymmetry $a_{sym}$ was observed and fitted for the maximal mass (MMS) neutron star scenario.  A general ordering in the strengths of these properties variation has been also obtained for the maximum mass star:
$$ \Delta M_{max}(\delta m_L) \overset{10 \times}{ >} \Delta M_{max}(\delta K) \overset{10 \times}{ >}  \Delta M_{max}(\delta a_{sym}) . $$  

Applying our phenomenological linear formulae we could determine precisely the Landau effective mass value, $m_{L} \approx 780$~MeV, which was consistent with our preliminary Bayesian analysis in Ref.~\cite{Alwarez:2020} as well.
Here we investigate the subsequent parameter in the above order, the compressibility $K$, which certainly has less impact on the maximal mass, than the radius, $R$~\cite{Chamel:2019,Kubis:2012,Klahn:2006,Margueron:2018}. Using our linear fits we present the parameter domain structure, while taking mass observation data from PSR J0740$+$6620~\cite{J0740}, PSR J0348$+$0432~\cite{J0348}, and PSR J1614$-$2230~\cite{J1614}, we estimate the mean $K$ value including the uncertainties originating from both the theory and the data.

\section{The model and the equation of state}
\label{sec:eos}

One of the most common description of the compact star's interior is the $\sigma $-$\omega$ model, which can be extended by further interactions~\cite{Posfay:2020,Posfay:2019}. This model describes protons, electrons, and neutrons in $\beta$-equilibrium and approximates the nuclear force by introducing the $\sigma$, $\omega$, and $\rho$ meson with higher-order self-interaction terms for the scalar meson. This model is the simplest one, which is able to describe the measured pulsars' mass and radius values, however more sophisticated descriptions are also available. In this study we focus on the connection between macro- and microscopical parameters, which led us to estimate nuclear parameters with high precision from pulsar data. 

The investigated Lagrange-function corresponding to the extended $\sigma$-$\omega$ model is, 
%
\begin{eqnarray}
\Lagr &=&
%
 \overline{\Psi} \left(
i \slashed{\partial} -m_{N} + g_{\sigma} \sigma -g_{\omega} \slashed{\omega}  + g_{\rho} \slashed{\rho}^{a} \tau_{a}
 \right) \Psi
+ \overline{\Psi}_{e} \left(
i \slashed{\partial} - m_{e}
\right) \Psi_{e}  \\
%
&& +\frac{1}{2}\,\sigma \left(\partial^{2}-m_{\sigma}^2 \right) \sigma - U_{i}(\sigma) 
%
- \frac{1}{4}\,\omega_{\mu \nu} \omega^{\mu\nu}+\frac{1}{2}m_{\omega}^2 \, \omega^{\mu}\omega_{\mu} 
%
-\frac{1}{4} \rho_{\mu \nu}^{a} \, \rho^{\mu \nu \, a} + \frac{1}{2} m_{\rho}^2 \, \rho_{\mu}^{a} \, \rho^{\mu \, a} \nonumber
\label{eq:wal_lag}
\end{eqnarray}
%
where $\Psi=(\Psi_{n},\Psi_{p})$ is the vector of proton and neutron fields, $m_{N}$ $m_{\sigma}$ $m_{\omega}$ are the masses of the nucleons and $\sigma$ and $\omega$ mesons, respectively. Furthermore  $g_{\sigma}$, $g_{\omega}$, and $g_{\rho}$ are the Yukawa couplings corresponding to the $\sigma$-nucleon,  $\omega$-nucleon, and $\rho$-nucleon interactions, respectively. The kinetic terms corresponding to the $\omega$ and $\rho$ meson are written as,
%
\begin{equation}
\omega_{\mu \nu}=\partial_{\mu} \omega_{\nu}-\partial_{\nu} \omega_{\mu}  \ \ \textrm{and } \ \
\rho_{\mu \nu}^{a}=\partial_{\mu} \rho_{\nu}^{a} - \partial_{\nu} \rho_{\mu}^{a} + g_{rho} \epsilon^{abc} \rho_{\mu}^{b} \rho_{\nu}^{c} \, .
\label{eq:wal_lag}
\end{equation}
%
In eq.~\eqref{eq:wal_lag}, $U_{i}(\sigma)$ is a self interaction term for the $\sigma$-meson and it has the following parametrization in this study:
%
\begin{equation}
U_{34}(\sigma) =\lambda_{3} \sigma^{3} + \lambda_{4} \sigma^{4}  \, .
\label{eq:U_types}
\end{equation}
%

We considered this model in the mean-field approximation at zero temperature and finite chemical potential. These assumptions simplify eq.~\eqref{eq:wal_lag}, where components of the mesons has non-zero value: $\omega_{0}=\omega$ and $\rho_{0}^{3}=\rho$, but kinetic terms are disappeared. From this point the free energy corresponding to the model can be calculated as it is described in for example in Ref.~\cite{jakovac2015resummation}: 
%
\begin{eqnarray}
f_{T} &=&
%
 f_{F}
\left(
m_{N}-g_{\sigma} \sigma,
\mu_{p} - g_{\omega} \omega + g_{\rho} \rho
\right)
+  f_{F} \left(
m_{N}-g_{\sigma} \sigma,
\mu_{n} - g_{\omega} \omega - g_{\rho} \rho
\right)
+ f_{F} \left(m_{e}, \mu_{e} \right)  \nonumber \\ 
%
&&+ \frac{1}{2} m_{\sigma}^{2} \sigma^2  + U_{i}(\sigma)
%
 - \frac{1}{2} m_{\omega}^2 \omega^2 
%
 - \frac{1}{2} m_{\rho}^2 \rho^2 \, , 
\label{eq:wal_f}
\end{eqnarray}
%
where $\mu_{p}$, $\mu_{n}$ and $\mu_{e}$ are the proton, neutron, and electron chemical potential, respectively. The $f_{F}$ term describes the free energy contribution corresponding to one fermionic degree of freedom, as usual 
%
\begin{equation}
f_{F}(T,m,\mu)  = -2 T \int \frac{\dd^3 k}{(2 \pi)^3} 
\ln{\left[ 1 + \mathrm{e}^{-\beta \left( E_{k}-\mu \right) } \right]}  \ \ \, 
\end{equation}
%
where $E_{k}^2  = k^2 + m^2$. In the cold, $T \to 0$ approximation, the description of the dense nuclear matter of the compact star, means that the fermionic free energy has only two variables $f_{F}(m, \mu)$. The free parameters of the model are determined by using nuclear saturation data~\cite{norman1997compact,meng2016relativistic}. The values used to fit the model are the binding energy $B=-16.3$ MeV, the saturation density, $n_{0} = 0.156$ fm\textsuperscript{-3}, the nucleon effective mass, $m^{*}= 0.6m_{N}$, the nucleon Landau mass $m_{L} =0.83 \, m_{N}$, together with compressibility and asymmetry energy $K =240$ MeV and $a_{sym} =32.5$ MeV, respectively. 
Following Ref.~\cite{norman1997compact}, the Landau mass is defined as, 
%
\begin{equation}
m_{L} =\frac{k_{F}}{v_{F}}  \quad \textrm{with} \quad
v_{F} =\left.\frac{\partial E_{k}}{\partial k} \right|_{k=k_{F}} \, .
\label{eq:landau_mass}
\end{equation}
%
Where $k=k_{F}$ the Fermi-surface and $E_{k}$ is the dispersion relation of the nucleons. The Landau mass is closely related to the effective nucleon mass in mean field theories: 
%
\begin{equation}
m_{L}= \sqrt{k_{F}^2 + m_{N, eff}^2} \, .
\label{eq:effmass_vs_landau_mass}
\end{equation}
%
The connection between the Landau mass and the nucleon effective mass can not let us to fit simultaneously both~\cite{meng2016relativistic}.
In this paper first the model is fitted to reproduce the data given above, except for the Landau and the effective nucleon mass. Then after getting the optimal Landau mass value, we estimate the values for the compressibility as well. 

These parameters are kept free and determined by comparing the mass radius diagrams corresponding to different values of the Landau mass to neutron star observations.  The compression modulus of the nuclear matter is defined as in Refs.~\cite{norman1997compact,Schmitt:2010}:
%
\begin{equation}
K =k_{F}^2 \frac{\partial^2 }{\partial k_{F}^2} \left( \frac{\epsilon}{n} \right)
= 9 n^2 \frac{\partial^2}{\partial n^2} \left( \frac{\epsilon}{n} \right) \, .
\label{eq:K}
\end{equation}
%
The asymmetry energy term originates from that nuclear force acting differently between proton and neutron states, and we can define it as,  
%
\begin{equation}
a_{sym} = \frac{1}{2} \left. 
\frac{\partial^2 }{\partial t^2} \left( \frac{\epsilon}{n} \right) \right|_{t=0}
\label{eq:asym_def}
\end{equation}
%
where $t=\frac{n_{n}-n_{p}}{n_{B}}$ is the relative fraction of the proton and neutron degrees of freedom. It's value is fitted as it is described for example in Ref.~\cite{norman1997compact}, but as we pointed out in Ref.~\cite{Posfay:2020} the value of the symmetry energy plays negligible role for maximal mass stars as it is well known from Refs.~\cite{Chamel:2019,Kubis:2012,Klahn:2006,Margueron:2018}.

For the general relativistic description of the compact stars we assumed the usual static picture in spherically symmetric space-time~\cite{norman1997compact,Haensel_book}. We calculated the mass-radius diagram by the Tolman\,--\,Oppenheimer\,--\,Volkoff equations (TOV)~\cite{Tolman:1939jz,Oppenheimer:1939ne},
%
\begin{equation}
\begin{aligned}
& \frac{\mathrm{d} p(r)}{\mathrm{d} r} =- \frac{G \epsilon(r) m(r)}{r^2} \times  \left[ 
1 + \frac{ p(r)}{\epsilon(r)}
\right]
\left[
1 + \frac{4 \pi r ^3 p(r)}{m(r)}
\right] 
 \left[
1 - \frac{2 G m(r)}{r}
\right]^{-1} \\[10pt]
 & \frac{\mathrm{d} m(r)}{ \mathrm{d} r } =4 \pi r^2 \epsilon(r)
\end{aligned}
\label{eq:TOV}
\end{equation}
%
where $p(r)$  and $\epsilon(r)$ are the pressure and energy density as functions of the radius of the star, $G$ is the gravitational constant while $m(r)$ is the mass of the star which is included in the mass shells up to the radius, $r$. To integrate the equations one need a connection between $p(r)$ and $\epsilon(r)$ at given $r$, which is provided by the equation of state (EoS) in the form of the relation $p(r)=p(\epsilon(r))$. To start the integration one has to choose a central energy density value, $\epsilon_{c}$ for the star as an initial condition.

To focus our investigation on the effect of the nuclear matter we integrated the TOV equations in the following way. Normally the integration is stopped when $p(r=R)=0$. However to get the correct result the EoS used to describe the neutron star at high densities is complemented by a low density EoS which describes the neutron star's crust. This introduces further parameters into the model description hence makes it harder to separate whether the observed effect can be attributed to the high density nuclear matter parameters or the low density EoS.

To circumvent this we employed a different stopping condition for the integration based on recent Ref.~\cite{Posfay:2020}, $p(r=R')=p_{0}$. Here $p_{0}$ is chosen such a way that the integration stops at the core of the neutron star, so the integration does not take into account the effect of the crust and the low density EoS. To get a good approximation for $p_{0}$ we used the well known BPS nuclear equation of state which is used to describe the crust of neutron stars~\cite{norman1997compact,BPS}. We used the highest pressure value for $p_{0}$ where the BPS EoS can still be considered valid.

The calculated $R'$ in this case corresponds to the radius of the neutron star core. The mass and radius data calculated this generally can be considered a conservative approximation of the neutron star parameters, however in the case of the maximum mass star the deviations from the normal case are insignificant \cite{Posfay:2020}.

\section{Connecting macroscopic and microscopic parameters}
\label{sec:fit}

Recent observations of precise maximal-mass pulsar data led us to characterize the effect of the most relevant parameter of the extended $\sigma$-$\omega$ model, the Landau mass, $m_L$. As a function of this microscopical nuclear property, we calculated the $M$-$R$ diagram and determined the mass and radius of the maximal mass stars (MMS). In this case we used the {\sl core approximation} from Ref.~\cite{Posfay:2020}. Assuming maximal mass stars, we found linear dependence of both $M_{maxM}$ and $R_{maxM}$ on the $m_L$, given by the independent, one-parameter formulae for asymmetric nuclear matter\footnote{We note, in Ref.~\cite{Posfay:2019} symmetric matter were investigated with similar results.}. In those cases where the Landau mass was optimized, we used the values from the saturated nuclear matter for the further nuclear parameters, $K$ and $a_{sym}$. 
Linear formulae for MMS stars' maximal mass and radius as a function of $m_L$ were fitted independently with 0.8\% and 17\% theoretical uncertainty, respectively, following Ref.~\cite{Posfay:2020}: 
\begin{align}
M_{maxM}(m_L)[\textrm{M}_{\odot}] = 5.418 - 0.00434 \, m_L [\textrm{MeV}] \ , \label{eq:maxmass_M-lm} \\ 
R_{maxM}(m_L)[\textrm{km}] = 19.04 - 0.01040 \, m_L [\textrm{MeV}]   \ . \label{eq:maxmass_R-lm} 
\end{align}

We found, the variation of the Landau mass values, $\delta m_L$ generates about an order of magnitude larger effect on the macroscopical observables, than tuning the compressibility, $\delta K$. Thus the compressibility plays a second-order role in the mass and radius parameters of the MMS class compact objects, similarly as the role of compressibility appears in measurements of high-energy nuclear collisions~\cite{Tsang:2008fd}. Since the phenomenological linear fits have about 10\% uncertainty, and the mean Landau mass was consistent for all the pulsars considered as MMS, one can try to make further fits for the compressibility values as well. After fixing the $m_L$ by the MMS observations we can also obtain linear, one-parameter dependence on the parameter $K$ with theoretical uncertainties, $\lesssim 2$~\% and  $\lesssim 14$~\%, respectively,
\begin{align}
M_{maxM}[\textrm{M}_{\odot}] = 1.766 + 0.00110 K \, [\textrm{MeV}] \ ,  \label{eq:maxmass_M-K} \\ 
R_{maxM}[\textrm{km}] = 8.878 + 0.00767 K \, [\textrm{MeV}] \ . \label{eq:maxmass_R-K} 
\end{align}
Dependence on the compressibility in eqs.~\eqref{eq:maxmass_M-K}-\eqref{eq:maxmass_R-K} have slope values with positive trends, which works against the negative slope values of the Landau mass dependence in eqs.~\eqref{eq:maxmass_M-lm}-\eqref{eq:maxmass_R-lm}. This means by increasing the compressibility, makes an MMS compact star more massive and larger -- as the equation of state gets softer. Finally, the effect of the variation of the $K$ is about 7 times stronger for the radius of the maximal mass star, than for the mass.

Using the astrophysical observation data so far, and assuming maximum mass pulsars, such as: 
PSR J0740$+$6620~\cite{J0740}, PSR J0348$+$0432~\cite{J0348}, and PSR J1614$-$2230~\cite{J1614}, we obtained the optimal Landau mass values for each pulsar in Table~\ref{tab:maxRM-lmk} and with the average of $m_{L} = 776.0^{+38.5}_{-84.9}$~MeV, similarly as in Ref.~\cite{Posfay:2020}. Furthermore, the compressibility can be also given by the independent formulae~\eqref{eq:maxmass_M-K}-\eqref{eq:maxmass_R-K}, thus we can use them to extend Table~\ref{tab:maxRM-lmk} with the compressibility values and the uncertainties indeed. For this case we calculated the mean compressibility value of $K=242.7^{+57.2 }_{-28.0}$~MeV, which agrees well with the saturated nuclear matter data. The radius of the maximal-mass neutron star is also calculated form eq.~\eqref{eq:maxmass_R-lm}. One can also obtain radii of the maximal mass stars form eq.~\eqref{eq:maxmass_R-K}. Both estimates overlap within the errorbars, but the latter one is more affected by the uncertainties.

\begin{table}[h!]
\begin{center}
\begin{tabular}{llccc}
\hline 
 Pulsar & $R_{maxM}$[km] & $M_{maxM}$[M$_{\odot}$] &$m_L$[MeV] &$K$[MeV]  \\
\hline
\hline
 PSR J0740+6620 & 11.25$^{+1.06}_{-1.04}$         & 2.17$^{+0.11  \ \ast}_{-0.10}$ & 748.39$^{+63.3 }_{-57.2}$  &  351.8$^{+115 }_{-84.5}$ \\
 PSR J0348+0432 & 10.87$^{+0.82}_{-0.80}$         & 2.01$^{+0.04  \ \ast}_{-0.04}$ & 785.25$^{+20.0 }_{-20.3}$ &  206.4$^{+42.7 }_{-20.5}$\\
 PSR J1614$-$2230 & 10.77$^{+0.82}_{-0.80}$       & 1.97$^{+0.04  \ \ast}_{-0.04}$ & 794.47$^{+20.1 }_{-20.4}$ &  170.0$^{+15.5 }_{-20.9}$ \\
\hline
\hline
\end{tabular}
\end{center}
\caption{\label{tab:maxRM-lmk} The Landau mass, $m_L$ and compressiblity, $K$ values calculated via eqs.~\eqref{eq:maxmass_M-lm} and~\eqref{eq:maxmass_M-K} from measured pulsar mass data denoted by '$\ast$', and assuming that these are maximal-mass neutron stars. The radii of these MMS stars are calculated by eq.~\eqref{eq:maxmass_R-lm}.}
\end{table}

\begin{figure}[!h]
\begin{center}
\includegraphics[width=0.92\textwidth]{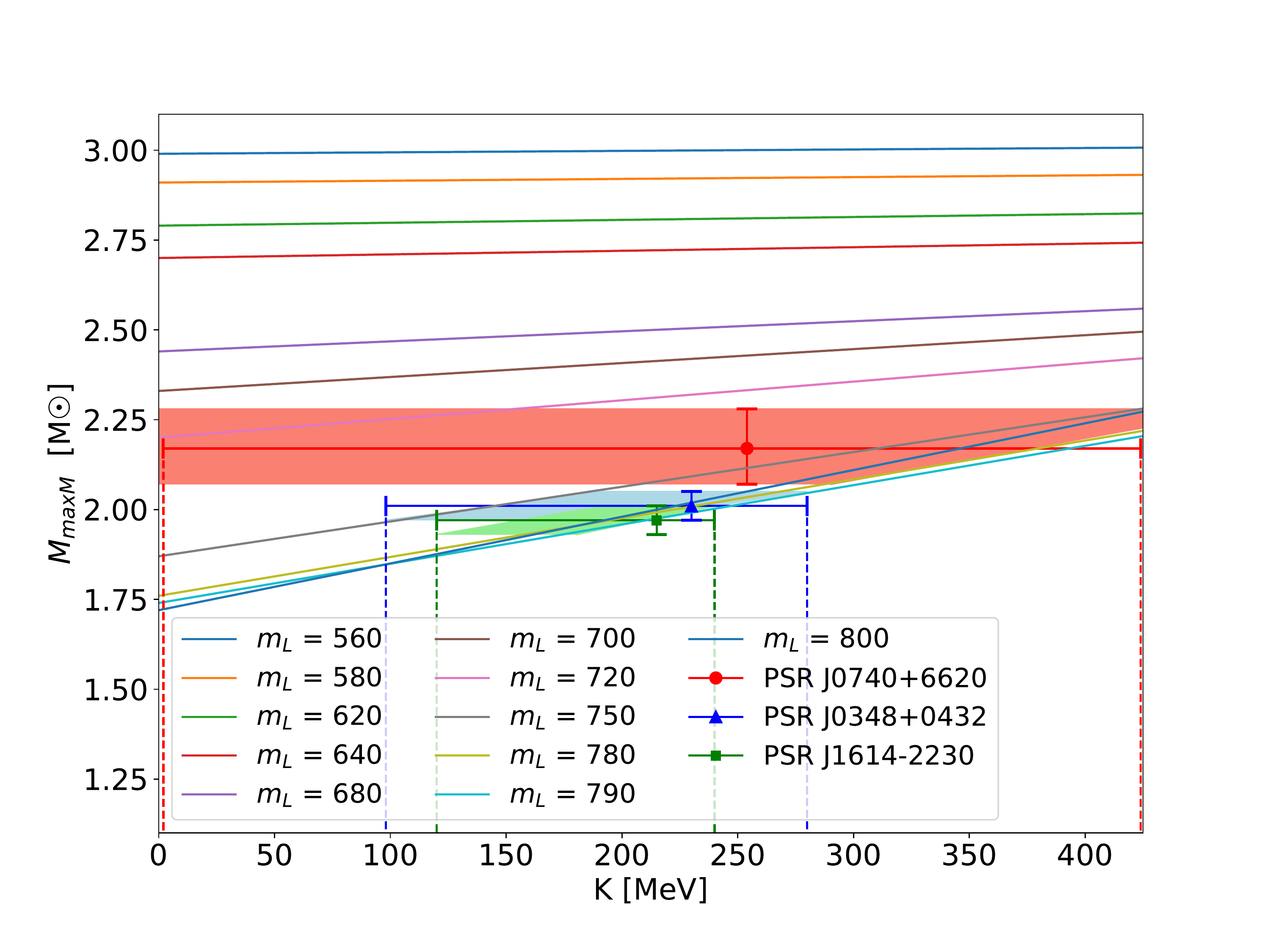} 
\includegraphics[width=0.92\textwidth]{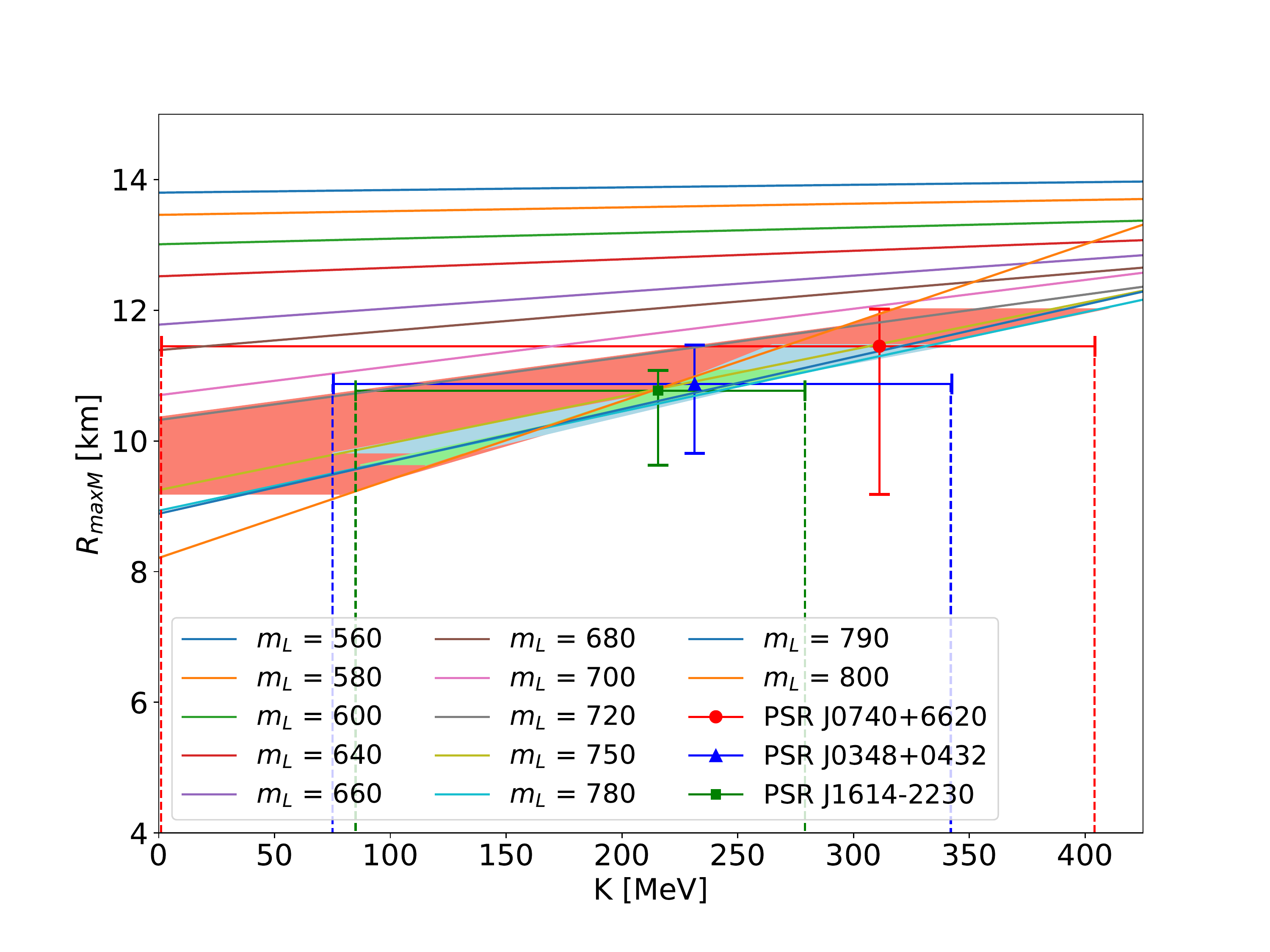} 
\end{center}
\caption{\label{fig:maxMass} Mass (top panel) and radius (bottom panel) of the maximal mass star. Curves are drawn as function of $K$ at various physical $m_L$ values, denoted as color lines. Comparisons to pulsar data of PSR J0740$+$6620~\cite{J0740}, PSR J0348$+$0432~\cite{J0348}, and PSR J1614$-$2230~\cite{J1614} are also plotted.}
\end{figure}

On Fig.~\ref{fig:maxMass} we plotted the evolution of the maximal mass and its radius dependence for the case of maximum mass stars (MMS). On the {\sl top panel} the projections of $M_{maxM}(m_L,K)$ are presented, where each function-line of $K$ was calculated for a fixed $m_L$ value. At the lowest $m_L=560$~MeV value the $K$-dependence is almost constant and it provides a 3~M$_{\odot}$ star independently of the $K$ values. As getting the highest value at $m_L=800$~MeV, lines are grouped with the same slope, but with lower offset. Pulsar mass data of PSR J0740$+$6620~\cite{J0740}, PSR J0348$+$0432~\cite{J0348}, and PSR J1614$-$2230~\cite{J1614} are indicated on Fig.~\ref{fig:maxMass} with color markers.  Uncertainties are plotted as errorbars and color-shaded areas, which include both  errors from the observations data and from the phenomenological fits.

Similar plot is shown on the {\sl bottom panel} of Fig.~\ref{fig:maxMass} for the $R_{maxM}(m_L,K)$. This presents steeper slope for the MMS star radii but with similar trends as on the {\sl top panel} the $M_{maxM}(m_L,K)$ curves. Radius of the MMS star at the lowest $m_L=560$~MeV value results almost constant, $R_{maxM} \approx 14$~km star for any $K$ values. While towards to the highest $m_L=800$~MeV the offsets decrease and the slopes increase, and finally this provides a lower boundary with a turning point in the $(m_L,K)$ parameter space. In addition to this, the calculated MMS pulsar radius data from eqs.~\eqref{eq:maxmass_M-mlK} and~\eqref{eq:maxmass_R-mlK} is also plotted. Note, errorbars and shaded areas were calculated from the theoretical uncertainties and observational data constraints. 

The above independent linear fits were expanded around the saturated nuclear matter values, but we can further improve these phenomenological formulae and present fits in the wider nuclear parameter domain used on Fig.~\ref{fig:maxMass}. The range of 550~MeV $< m_L <800$ MeV is compatible with effective nucleon mass given by the various equation of states at saturation~\cite{Klahn:2006}, while for the compressibility interval, $K <450$ MeV was chosen for our investigation. Within this wide domain the entanglement of the $m_L$ and $K$ parameters presents. Dependence on these variables are not independent any more, but can be factorized into joint formulae, including higher-order, cross-product terms for the $M_{maxM}(m_L,K)$ and $R_{maxM}(m_L,K)$ expressions, respectively:  
\begin{equation}
\resizebox{0.920\hsize}{!}{
$M_{maxM}[\textrm{M}_{\odot}] = 6.29 - 0.00574\, m_L\, [\textrm{MeV}] - 0.00379\, K\, [\textrm{MeV}] + 0.00000524 \, m_L \cdot K\, [\textrm{MeV}^2] \, , $ }
\label{eq:maxmass_M-mlK}
\end{equation}
\begin{equation}
\resizebox{0.920\hsize}{!}{
$R_{maxM} [\textrm{km}] =
27.51 - 0.0239\, m_L\, [\textrm{MeV}]- 0.0241\, K\, [\textrm{MeV}] + 0.0000411\, m_L \cdot K\, [\textrm{MeV}^2]  \, . $} 
\label{eq:maxmass_R-mlK}
\end{equation}

These phenomenological expressions were expanded around the averaged parameter values obtained by the constraints from the astronomical MMS pulsar data observations. Considering eqs.~\eqref{eq:maxmass_M-mlK} and~\eqref{eq:maxmass_R-mlK}, one can see that the offset values for both nuclear parameters got larger. On the contrary, all the linear slopes become negative, which are compensated by positive cross-product terms of the Landau mass and the compressibility, $ m_L \cdot K$. 

We note, taking expressions~\eqref{eq:maxmass_M-mlK} and~\eqref{eq:maxmass_R-mlK} at their averaged parameter values, one can get back the original independent, one-parameter equations, however these have slightly different values and larger uncertainties of about 20\%. This difference reflects the consequences of the '{\sl linearization}' of the complex and entangled nuclear parameter dependence in the physical-relevant domain -- as it was presented in Ref.~\cite{Alvarez:2020b} as well as the limited observational constraints.

\section{Discussion of the results}
\label{sec:res}

To explore the proper cross-dependence between $m_L$ and $K$, we investigated the mass and radius configurations of compact stars within the extended $\sigma$-$\omega$ model and using a wide survey on the parameter values. We used the interval 550~MeV$\le m_L\le $800~MeV and we mapped the configurations for the $K\le 450$~MeV values. The analysis was restricted to maximal mass compact stars (MMS) only, where the core approximation can be applied. For the parameter analysis we calculated maximal mass and the corresponding radius values as the function of the $m_L$ and $K$. We found that the variation of the compressibility modifies the Landau mass values as well, and this works {\sl vice versa}. This appears as 5-10\% variation in the values of slopes and the offset parameters of the maximal mass and the corresponding radius formulae. It is interesting to see, that in the linear approximations, a natural limit or turning point is present for the $M_{maxM}(m_L,K)$ and $R_{maxM}(m_L,K)$ expressions within the physically-relevant parameter domain, around the highest $m_L$ values. 
One can identify the 'masquarade problem' as an induced uncertainty of this dense parameter domain transmitted via the Tolman\,--\,Oppenheimer\,--\,Volkoff equations to macroscopic observables. 

\begin{figure}[!h]
\begin{center}
\includegraphics[width=0.92\textwidth]{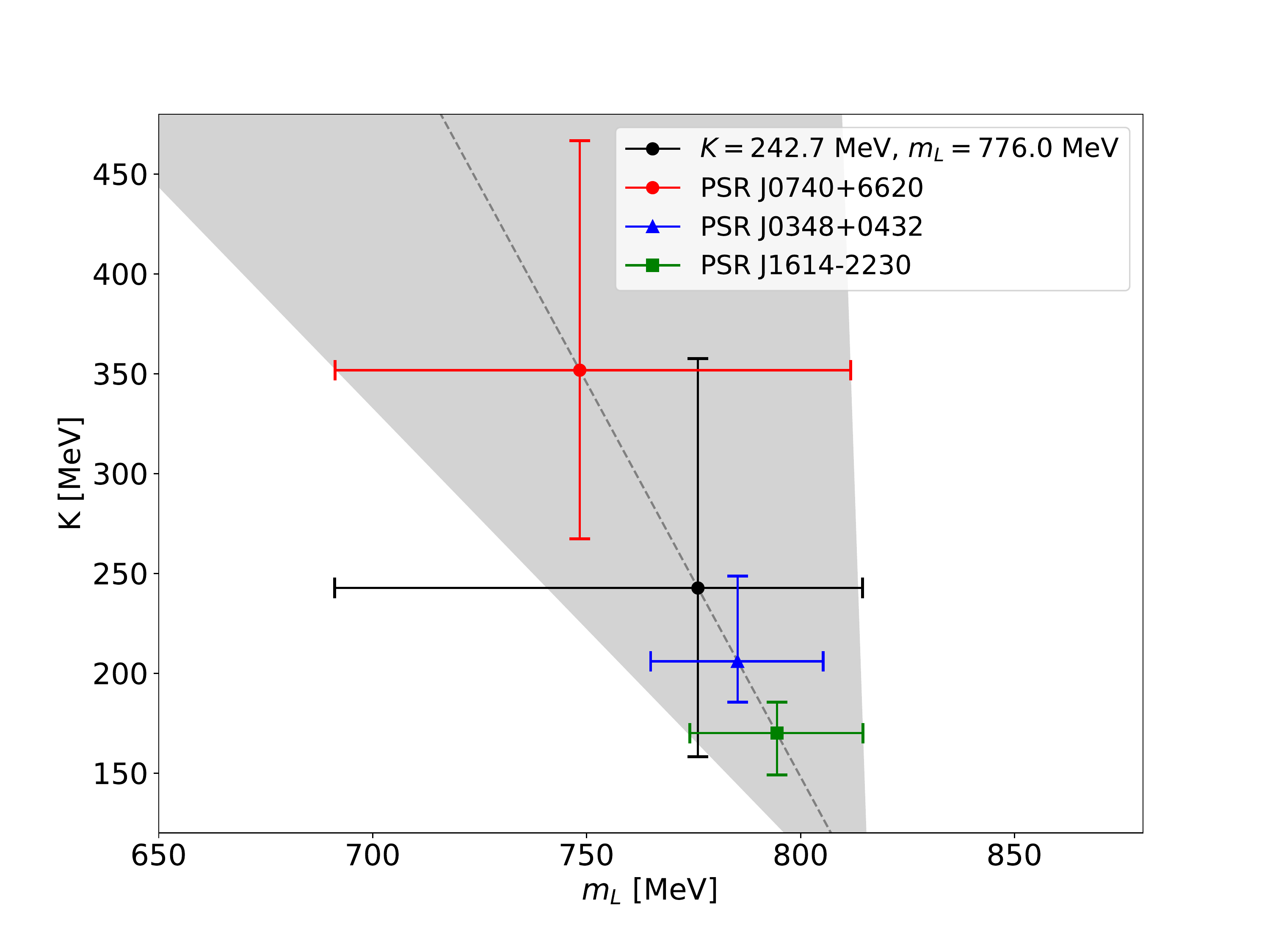} 
\end{center}
\caption{\label{fig:NuclParam} The nuclear model parameter space of $K$ and $m_L$ is plotted with values estimated from pulsar data of PSR J0740$+$6620~\cite{J0740}, PSR J0348$+$0432~\cite{J0348}, and PSR J1614$-$2230~\cite{J1614}. Fit line and average microscopic data value with uncertainty is also marked.}
\end{figure}

On Figure~\ref{fig:NuclParam} the microscopical nuclear parameter domain of the Landau mass and the compressibility is plotted in order to summarize and compare the obtained microscopical nuclear parameter values. {\sl Color markers} are for the pulsar data and the {\sl black point} is for the average value calculated from the astrophysical data constraints. We indicated with {\sl dashed line} the interconnected data points used in the expansions above with  a conservative uncertainty estimate ({\sl shaded area}).

Thanks to the maximal mass star assumption we obtained relatively-precise phenomenological formulae, thus the average of the Landau mass values and the compressiblity can be estimated $m_{L} = 776.0^{+38.5}_{-84.9}$~MeV and $K=242.7^{+57.2 }_{-28.0}$~MeV, respectively. These results are well-compatible with the saturated nuclear matter parameters, and also overlaps with the parameter domain of various equation of states as summarized in Ref.~\cite{Klahn:2006}. Within these mean field models the effective nucleon mass is typically $0.54m_N<m^{\ast}<0.8 m_N$, therefore our average values overlap with the range of the Landau mass, assuming $m_L \approx 1.4m^{\ast}$. Furthermore, the predicted compressibility values are within the range of 203~MeV~$<K<275$~MeV, accordingly~\cite{Klahn:2006}. Moreover, laboratory measurements of $K\approx 240 \pm 20$~MeV also support our result~\cite{Shlomo:2006,Piekarewicz:2010,Li:2013,Oertel:2017}.

As a cross-check, our recent phenomenological results from the linear approximations can be also compared to a more general Bayesian analysis using the same extended $\sigma$-$\omega$ model based calculation. In our early work, we have estimated $m_L = 750\pm 15$~MeV by a complex method, however in this study we varied only one parameter, while keeping the others fixed with values of $n_0 = 0.156$~fm\textsuperscript{-3}, $K = 240$~MeV and $a_{sym} = 32.5$~MeV~\cite{Alwarez:2020}. In the improved continuation of this investigation, we used a set of 10\textsuperscript{3} equation of states with the combinations of the nuclear parameter values of $m_L$, $K$, and $a_{sym}$. In this analysis the applied scenarios were constrained by gravitational wave measurements and measured compact star radii data as well, and certainly here crust was also included. The obtained parameters are: $m_L =727.4\pm 15$~MeV, $K=232\pm 20$~MeV, and $a_{sym}=31.8\pm 3.8$~MeV~\cite{Alvarez:2020b}. 

\section{Summary}
\label{sec:sum}

We investigated the consequences of precise pulsar mass measurements in relation with the extended $\sigma$-$\omega$ mean field model of the cold super-dense nuclear matter. We selected pulsar mass data: PSR J0740$+$6620~\cite{J0740}, PSR J0348$+$0432~\cite{J0348}, and PSR J1614$-$2230~\cite{J1614} assuming that these are maximal mass stars (MMS), for which the {\sl core approximation} from Ref.~\cite{Posfay:2020} is suitable. By this assumption: observational parameters of these MMS are mainly determined by the core of the star and crust plays negligible role on the mass and radius of the star around the highest mass values. 

We have found that the microscopical nuclear parameters: Landau mass, $m_L$ and nuclear compressibility $K$, determine well the mass and radius of the MMS around this point. Since the extended $\sigma$-$\omega$ model describes well the core of a compact object with limited number of parameters, we obtained linear relations between the microscopic and macroscopic parameters of maximal mass stars within 10\% precision. 

Expanding our formulae around these parameters, results in two-parameter functions for the $M_{maxM}(m_L,K)$ and $R_{maxM}(m_L,K)$. We investigated the relations between microscopic parameters and macroscopical observables. Although the structure of the microscopical parameter domain suggested the presence of the {\sl masquarade problem}, combining pulsar data with the phenomenological formulae, led us to determine the value of the nucleon Landau mass and the nuclear compressibility, $m_{L} = 776.0^{+38.5}_{-84.9}$~MeV and $K=242.7^{+57.2 }_{-28.0}$~MeV, respectively. Moreover the missing radius parameters for MMS were also obtained with $R=10.96^{+1.35}_{-1.00}$.

However, our obtained results are specified for MMS stars only, but we have found them consistent with a more complex study applying Bayesian algorithm for the determination of the microscopical nuclear matter parameters in Refs.~\cite{Alwarez:2020,Alvarez:2020b}. Within the framework of this more general method it is allowed to use not only maximal mass star data, but other arbitrary mass and radius measurements. Since both the obtained microscopical parameter values and the predicted observational data are consistent if constraints are the same, this supports well the applicability of our phenomenological linear formulae of the present investigation.


\section*{Acknowledgements}
Author GGB acknowledges the fruitful discussions with David Blaschke, his infinite energy, criticisms, and friendship provided great motivation for me. I wish you David the full-bull-power for the forthcoming many-many years for your birthday(s)!

This work is supported by the Hungarian Research Fund NKFIH (OTKA) under contracts No. K120660, K123815, and COST actions PHAROS (CA16214) and THOR (CA15213). Authors also acknowledge the computational resources for the Wigner GPU Laboratory.
~

\end{document}